\documentclass[]{aa}
\usepackage[dvips]{graphicx}
\usepackage{txfonts}
\begin{document}
\title{The $K$-band intensity profile of R\,Leonis probed by VLTI/VINCI 
\thanks{Based on public commissioning data released from the ESO VLTI
(www.eso.org/projects/vlti/instru/vinci/vinci\_data\_sets.html).}
}
\author{
D.~Fedele\inst{1,2} \and
M.~Wittkowski\inst{1} \and
F.~Paresce\inst{1} \and
M.~Scholz\inst{3,4} \and
P.~R.~Wood\inst{5} \and
S.~Ciroi\inst{2}
}
\offprints{M.~Wittkowski, \email{mwittkow@eso.org}}
\institute{
European Southern Observatory, Karl-Schwarzschild-Str. 2,
85748 Garching bei M\"unchen, Germany
\and
Dipartimento di Astronomia, Universit\`a di Padova, 
Vicolo dell'Osservatorio 2, 35122 Padova, Italy
\and
Institut f\"ur Theoretische Astrophysik der Universit\"at Heidelberg,
Albert-Ueberle-Str. 2, 69120 Heidelberg, Germany
\and
School of Physics, University of Sydney, NSW 2006, Australia
\and
Research School for Astronomy and Astrophysics, Australian National
University, Canberra, ACT 2600, Australia
}
\titlerunning{The $K$-band intensity profile of R\,Leo}
\date{Received \dots; accepted \dots}
\abstract{We present near-infrared $K$-band interferometric measurements 
of the Mira star R\,Leonis obtained in April 2001 and January 2002
with the VLTI, the commissioning instrument VINCI, 
and the two test siderostats. These epochs correspond to near-maximum
stellar variability phases $\sim$\,0.08 and $\sim$\,1.02 (one cycle later), 
respectively. The April 2001 data cover a range of spatial 
frequencies (31--35 cycles/arcsecond) within the first 
lobe of the visibility function. These measurements indicate 
a center-to-limb intensity variation (CLV) that is clearly different
from a uniform disk (UD) intensity profile. We show that
these measured visibility values are consistent with predictions from recent
self-excited dynamic Mira model atmospheres that 
include molecular shells close to continuum-forming layers. 
We derive high-precision Rosseland diameters 
of $28.5 \pm 0.4$\,mas and $26.2 \pm 0.8$\,mas for the April 2001 and 
January 2002 data, respectively. Together with literature estimates of the 
distance and the bolometric flux, these values
correspond to linear radii of $350^{+50}_{-40}$\,R$_\odot$ and 
$320^{+50}_{-40}$\,R$_\odot$, and to effective temperatures 
of $2930 \pm 270$\,K and $3080 \pm 310$\,K, respectively.

}
\maketitle
\section{Introduction}
\label{sec:introduction}
\object{R\,Leo} is one of the brightest, apparently largest, and hence
best studied Mira variable stars. Mira stars are cool, low-mass, pulsating 
variables that are located on the asymptotic giant branch of 
the Hertzsprung-Russell 
diagram and that exhibit a conspicuous mass-loss. Because of to the low 
temperatures, molecules are present in their extended atmospheres, and dust
is formed at larger distances from the star. 
High resolution techniques at optical/infrared wavelengths, such as
speckle interferometry, masked aperture techniques, lunar occultation 
measurements, and long-baseline interferometry, allow
the measurement of the stellar radii,
the center-to-limb intensity variations (CLV) including the effects
of molecular layers, and the dust envelopes.
For instance, Mira star diameters and CLVs have been measured by, e.g.,
Di Giacomo (\cite{digiacomo}), 
Haniff et al. (\cite{haniff}), Burns et al. \cite{burns},
Perrin et al. (\cite{perrin}), Young et al. (\cite{young}), 
Hofmann et al. (\cite{hofmann}), Hofmann et al. (\cite{hofmann02}),
Woodruff et al (\cite{woodruff}), Perrin et al. (\cite{perrin2004}).
Works by, e.g., Danchi et al. (\cite{danchi}), 
Bedding et al. (\cite{bedding}), 
Monnier et al. (\cite{monnier}), Ohnaka et al. (\cite{ohnaka}) also include
 characterizations of circumstellar dust shells.
For non-variable giants up to at least spectral type M\,4, 
it was shown that observed and model-predicted visibility curves 
do not differ significantly from those of uniform disk (UD) profiles up to 
the first minimum, but result in a diameter that is different by a few percent 
at near-infrared wavelengths (see, e.g the discussions in 
Wittkowski et al. \cite{WHJ:01,WAK:04}). For Mira stars, however,
it was observed that the visibility function may deviate already in its
first lobe from a UD curve, indicating a two- or multi-component,
or Gaussian-shaped intensity profile.
Theoretical models including the effects of molecular shells
have been created that can explain such CLVs. For more
detailed discussions on these topics, see the reviews by 
Scholz (\cite{Scholz98}, \cite{Scholz03}), and also Scholz (\cite{Scholz01}), 
as well as the references given above. 
In particular, the P and M self-excited dynamic atmosphere model series 
by Hofmann et al. (\cite{HSW98}), Tej et al. (\cite{tej03}), 
Ireland et al. (\cite{ISW,ISTW}), have been designed to match the 
well-studied nearby Miras $o$\,Cet and R\,Leo. 
These series are based on fundamental mode pulsation models. 
The comparison
of theoretical pulsation models with MACHO observations
of long-period variables in the LMC (Wood et al. \cite{wood}),
pulsation velocities derived from Doppler line profiles (Scholz \& 
Wood \cite{scholzwood}), as well as recent detailed analyses of
interferometric radius measurements
(for instance, Mennesson et al. \cite{mennesson}, 
Woodruff et al. \cite{woodruff}, Perrin et al. \cite{perrin2004})
strongly indicate that Mira stars are fundamental mode pulsators.

Earlier diameter estimates of Mira stars which suggested overtone
pulsation very likely overestimated the true photospheric diameter
due to the effects of molecular layers at larger distances from the star
(cf. for instance the discussions in 
Jacob \& Scholz \cite{JS02};
Mennesson et al. \cite{mennesson};
van Belle et al. \cite{vanbelle2};
Woodruff et al. \cite{woodruff};
Ireland et al. \cite{ISW,ISTW};
Perrin et al. \cite{perrin2004}). In addition, distances 
to Mira stars are often not known with a high precision since the parallax
is typically only a fraction of the stellar angular diameter. 
This leads to an additional uncertainty when comparing measured 
Mira star radii to theoretical radius-period relations.
Note, however, that the pulsation mode 
determination in Wood et al. (\cite{wood}) depends mainly on the 
period {\em ratios} in multiperiodic long-period variables. 
Since period ratios are unaffected by uncertainties in distance or 
angular diameter measurements, this mode determination should be robust.

Often, interferometric measurements cannot probe the CLV in detail
owing to sparse coverage of the $uv$-plane. In these cases,
filter-specific uniform disk radii are often transformed into
physically more meaningful Rosseland or continuum radii using
available Mira star atmosphere models 
(e.g. van Belle et al. \cite{vanbelle1,vanbelle2};
Hofmann et al. \cite{hofmann02}; Boboltz \& Wittkowski \cite{BW}).
Detailed tests of Mira star atmosphere models at different
stellar variability phases are desirable in order to increase
the confidence in using them for such purposes.
Recently, Woodruff et al. (\cite{woodruff}) have shown that their
near-infrared $K$-band interferometric observations of $o$\,Cet
indicate a CLV that differs from a UD profile, and that
is consistent with CLV predictions by the P and M model series 
mentioned above. 

In this paper, we present a comparison of 
near-infrared $K$-band VLTI/VINCI interferometric observations of R\,Leo
with predictions by the P and M model series.

R\,Leo is an oxygen-rich Mira star with spectral type 
M6e-M8IIIe-M9.5e, a period of 310 days, a $V$ magnitude of
4.4-11.3 (Kholopov et al. \cite{kholopov}), and a mass-loss rate
of $\sim$\,1\,$\times 10^{-7}$\,M$_\odot$/yr 
(Danchi et al. \cite{danchi}, Knapp et al. \cite{knapp}). 
Sloan \& Price (\cite{sloan})
measured a relatively low dust emission coefficient of 0.23, i.e. the ratio of 
the total
emission of the dust to the total emission of the star in the
mid-infrared. We use a parallax value 
of $8.81 \pm 1.00$\,mas as given by Whitelock \& Feast (\cite{WF}),
which is the  weighted average of the values by Gatewood (\cite{gatewood}) 
and Perryman \& ESA (\cite{esa}). 
Whitelock et al. (\cite{whitelock}) derived a mean bolometric magnitude 
$m_\mathrm{bol}$=0.65 with total (peak-to-peak) amplitude
$\Delta m_\mathrm{bol}$=0.63.
Previous measurements of the $K$-band diameter of R\,Leo are 
listed in Table~\ref{tab:diams}, together with the visual phases at the 
time of the observations.
\begin{table}
\centering
\caption{Previous measurements of R\,Leo's $K$-band UD diameter,
together with the visual variability phase at the time of observation.}
\begin{tabular}{cll}
\hline
\hline
$\Theta_\mathrm{UD}^K$[mas]   & $\phi_\mathrm{vis}$    & Reference    \\
\hline
33.10 $\pm$ 1.30 & 0.18  & Di Giacomo et al. (\cite{digiacomo}) \\
28.18 $\pm$ 0.05 & 0.24  & Perrin et al. (\cite{perrin}) \\
30.68 $\pm$ 0.05 & 1.28  & Perrin et al. (\cite{perrin}) \\
34.00 $\pm$ 2.00 & 0.44  & Tej et al. (\cite{tej99}) \\
30.00 $\pm$ 0.30 & 0.71  & Monnier et al. (\cite{monnier}) \\
\hline
\end{tabular}
\label{tab:diams}
\end{table}
\section{Observations and data reduction}
\begin{figure}
\resizebox{\hsize}{!}{\includegraphics{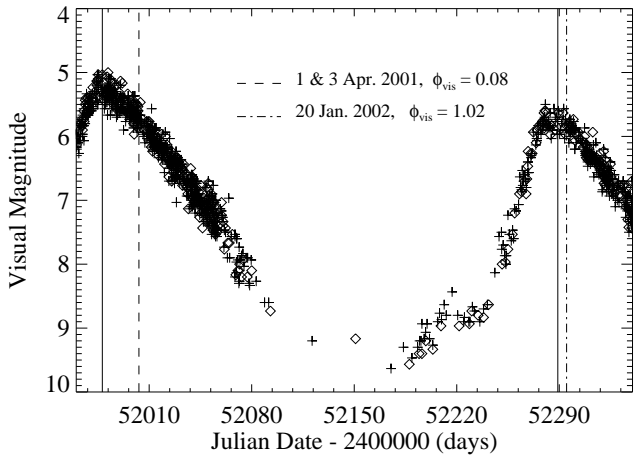}}
\caption{Visual lightcurve of R\,Leo for the cycles covering our
observations. The data are from the AAVSO ('+' symbols, Waagen \cite{AAVSO})
and AFOEV (diamond symbols, operated by CDS) databases. The solid lines 
denote the 
dates of maximum visual light as given by the AFOEV database. 
The dashed and dashed-dotted lines denote the dates of our observations.
Note the post-minimum ``bump'' (see text).}
\label{fig:lightcurve}
\end{figure}

\begin{table}
\centering
\caption{Details of our observations (date and time of observation,
spatial frequency, azimuth angle of the projected baseline (E of N)),
together with the measured squared visibility amplitudes and their
errors. The last column denotes the number of successfully
processed interferograms for each series. The effective
wavelength for R\,Leo is $\sim$\,2.22\,$\mu$m.}
\label{tab:obs}
\begin{tabular}{lccccl}
\hline\hline
 UT & Sp. freq  & az  & $V^2$  & $\sigma_{V^2}$ & \# \\
 &[1/$^{\prime\prime}$]& [deg] &   &  &   \\\hline 			 
\multicolumn{6}{c}{1 April 2001}\\
 02:25:26 & 34.93       &  72.86   & 5.07e-02 & 2.46e-03 &  389\\
 02:35:12 & 34.95       &  72.39   & 5.40e-02 & 2.86e-03 &  306\\
\multicolumn{6}{c}{3 April 2001}\\
 00:36:01 & 31.39       &  76.30   & 8.14e-02 & 4.08e-03 &  388\\
 00:40:13 & 31.66       &  76.21   & 8.13e-02 & 4.56e-03 &  332\\
 00:44:27 & 31.91       &  76.11   & 7.65e-02 & 3.76e-03 &  391\\
 00:49:06 & 32.18       &  76.00   & 7.55e-02 & 3.78e-03 &  374\\
 00:53:45 & 32.44       &  75.88   & 7.21e-02 & 3.58e-03 &  389\\
 00:58:31 & 32.69       &  75.76   & 7.10e-02 & 3.52e-03 &  383\\
 01:03:09 & 32.92       &  75.63   & 6.99e-02 & 3.48e-03 &  379\\
 01:08:14 & 33.17       &  75.48   & 6.37e-02 & 3.16e-03 &  386\\
 01:12:44 & 33.36       &  75.35   & 6.09e-02 & 3.01e-03 &  380\\
 01:17:27 & 33.56       &  75.21   & 6.07e-02 & 3.03e-03 &  379\\
 01:21:18 & 33.71       &  75.08   & 5.85e-02 & 3.20e-03 &  338\\
 01:25:35 & 33.87       &  74.94   & 5.63e-02 & 2.79e-03 &  387\\
 01:30:12 & 34.03       &  74.79   & 5.63e-02 & 2.79e-03 &  387\\
 01:34:37 & 34.17       &  74.63   & 5.06e-02 & 3.18e-03 &  146\\
 01:38:50 & 34.29       &  74.48   & 5.32e-02 & 2.58e-03 &  382\\
 01:42:36 & 34.40       &  74.34   & 5.42e-02 & 2.83e-03 &  371\\
 01:47:20 & 34.51       &  74.16   & 5.10e-02 & 2.46e-03 &  388\\
 01:51:09 & 34.60       &  74.01   & 5.15e-02 & 2.62e-03 &  386\\
 01:55:20 & 34.67       &  73.84   & 4.99e-02 & 2.41e-03 &  387\\
 01:59:10 & 34.74       &  73.68   & 4.95e-02 & 2.56e-03 &  376\\
 02:03:24 & 34.80       &  73.50   & 4.89e-02 & 2.36e-03 &  394\\
 02:07:14 & 34.85       &  73.33   & 4.88e-02 & 2.46e-03 &  383\\
 02:13:10 & 34.90       &  73.06   & 4.80e-02 & 2.31e-03 &  393\\
 02:16:57 & 34.93       &  72.89   & 4.85e-02 & 2.44e-03 &  387\\
 02:21:09 & 34.94       &  72.69   & 4.72e-02 & 2.27e-03 &  389\\
 02:42:02 & 34.88       &  71.62   & 4.85e-02 & 2.33e-03 &  391\\
 02:45:46 & 34.84       &  71.41   & 4.91e-02 & 2.50e-03 &  382\\
 02:50:36 & 34.78       &  71.14   & 4.96e-02 & 2.38e-03 &  391\\
 02:54:38 & 34.71       &  70.90   & 4.88e-02 & 2.45e-03 &  384\\
 02:58:43 & 34.64       &  70.66   & 5.02e-02 & 2.42e-03 &  269\\
\multicolumn{6}{c}{20 January 2002}\\
 06:22:56 & 34.18       &  74.62   & 8.73e-02 & 1.65e-03     & 453\\
 06:29:14 & 34.36       &  74.39   & 8.67e-02 & 4.45e-03     & 285\\
 07:20:33 & 34.94       &  72.12   & 7.93e-02 & 1.44e-03     & 487\\
 07:26:12 & 34.91       &  71.82   & 7.87e-02 & 1.47e-03     & 466\\
\hline
\end{tabular}
\end{table}
The R\,Leo interferometric data were obtained on 
1 \& 3 April 2001 (JD=2452003, stellar phase $\phi_\mathrm{vis}=0.08$) 
and on 20 January 2002 (JD=2452295, $\phi_\mathrm{vis}=1.02$) with
the ESO Very Large Telescope Interferometer (VLTI) equipped with the 
$K$-band commissioning instrument VINCI (Kervella et al. \cite{kervella1}).
The VLTI test siderostats were used on stations E0 and G0 forming an 
unprojected ground baseline length of 16\,m. 
These data were obtained in the early commissioning phase of the VLTI,
and are publicly available from the ESO archive.
Figure~\ref{fig:lightcurve} shows the visual lightcurve for the 
cycles covering our observation dates, as obtained from the AAVSO
(Waagen \cite{AAVSO}) and AFOEV (operated by CDS) databases. Indicated 
are the dates of our observations as well as the dates of maximum visual light,
as given by the AFOEV database for these cycles. It shows a cycle-to-cycle
variation from our first to our second epoch of $\sim$\,0.5\,mag near
the maxima. 
Note that the post-minimum ``bump''
which is predicted by self-excited pulsation models 
(Hofmann et al. \cite{HSW98}; Ireland et al. \cite{ISW,ISTW}) is 
clearly seen in Fig.~\ref{fig:lightcurve}.
Tab.~\ref{tab:obs} lists the details of our 
observations together with the obtained calibrated squared visibility 
amplitudes and their errors. 
Data were obtained as a series of, typically, 500 interferograms with
scan lengths of 250 $\mu$m and fringe frequencies of 295\,Hz and 415\,Hz.
Mean coherence factors were computed using the VINCI data reduction
software, version 3.0, as described by Kervella et al. (\cite{kervella2}),
employing the results based on the wavelets power spectral density.
The stars \object{$\alpha$\,Hya} ($K$-band UD diameter 
$\Theta_\mathrm{UD}^K=9.10 \pm 0.91$\,mas, using the calibration
by Dyck et al. \cite{dyck}), 
\object{$\delta$\,Oph} ($\Theta_\mathrm{UD}^K=9.76 \pm 0.10$\,mas, Bord\'e 
et al. \cite{borde}), \object{$\theta$\,Cen} 
($\Theta_\mathrm{UD}^K=5.32 \pm 0.06$\,mas, Bord\'e et al. \cite{borde}),
and \object{$\upsilon$\,Cet} ($\Theta_\mathrm{UD}^K=5.18 \pm 0.51$\,mas, 
after Dyck et al. \cite{dyck}) were used as calibration stars. 
The calibration of the visibility values
was performed as described in Wittkowski et al. (\cite{WAK:04}), using
a weighted average of all transfer function values obtained during the
night. R\,Leo data were calibrated using calibration star data obtained with
the same fringe frequency while the transfer function errors were computed
using all available values.
The listed errors of the calibrated squared visibility values include 
the uncertainties
of the adopted calibration stars' diameters, the variation of the 
transfer function over the night, as well as the scatter of the single
scans' coherence factors. For the night 3 April 2001, all calibration
star data were obtained after all R\,Leo observations. 
This may cause an additional systematic error of the calibrated visibility 
values which is not included in the listed error bars. However, it is 
very unlikely that such an additional calibration uncertainty 
is larger than the error bars shown.
All observations cover a narrow range of
azimuth angles of the projected baseline (70-76 deg. E of N), and hence are
not sensitive to possible asymmetries. The VINCI sensitivity
function covers the range 1.9 to 2.5\,$\mu$m. The effective wavelength
of our R\,Leo observations is $\lambda_0~\sim$\,2.22\,$\mu$m.

In order to derive effective temperatures from the measured angular radius
and the bolometric flux (see Sect.~\ref{sec:models} below), we use the
mean bolometric 
magnitude and its amplitude given by
Whitelock et al. (\cite{whitelock}). 
For the stellar phases 0.08 and 1.02 of our observations, we derive 
bolometric fluxes of
$f_\mathrm{bol}(\mathrm{phase}=0.08)
=(1.98 \pm 0.59) \times 10^{-5}$\,erg\,cm$^{-2}$\,s$^{-1}$, and 
$f_\mathrm{bol}(\mathrm{phase}=1.02)
=(2.05 \pm 0.61) \times 10^{-5}$\,erg\,cm$^{-2}$\,s$^{-1}$, respectively.
Errors of 30\% are considered, taking into account
cycle-to-cycle variations and the uncertainty of the 
estimate at phases 0.08 and 1.02.
\section{Results and comparison to models}
\label{sec:models}
\begin{figure}
\resizebox{\hsize}{!}{\includegraphics{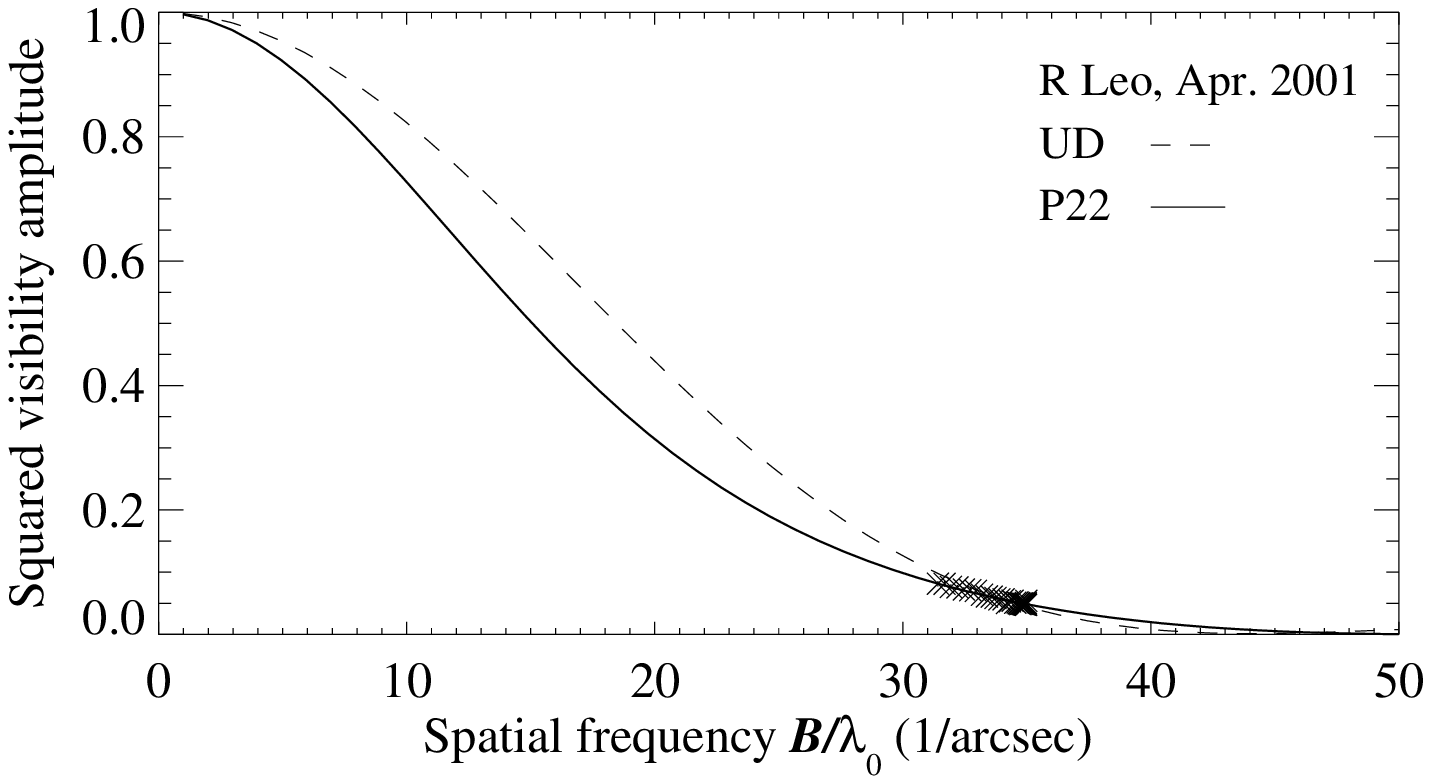}}

\resizebox{\hsize}{!}{\includegraphics{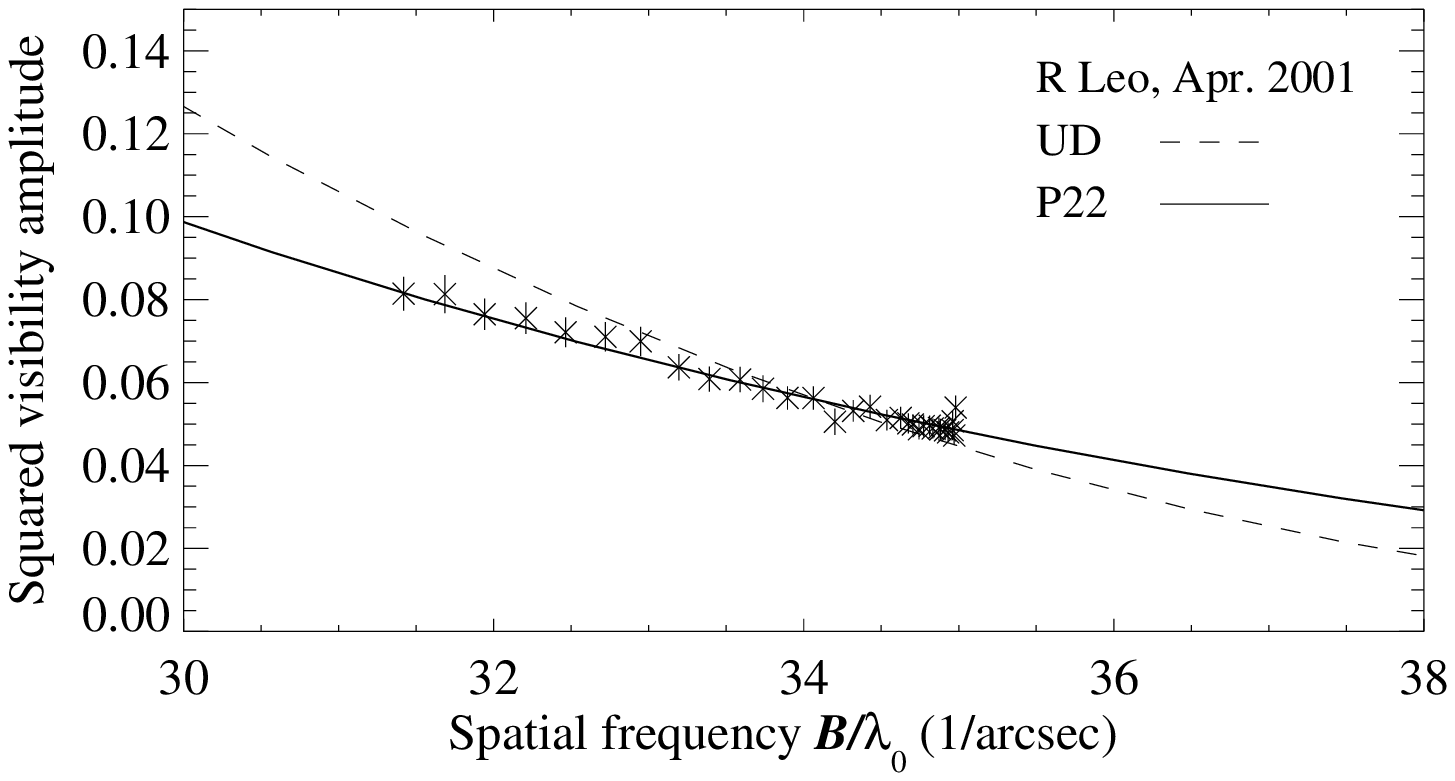}}
\caption{Measured R\,Leo squared visibility amplitudes obtained
in April 2001 (visual stellar phase 0.08), together
with the well fitting P22 model prediction. For comparison, the UD curve
is shown as well. The upper panel shows the whole range of
spatial frequencies while the lower panel shows an enlargement
of the measured values.}
\label{fig:vis2001}
\end{figure}

\begin{figure}
\resizebox{\hsize}{!}{\includegraphics{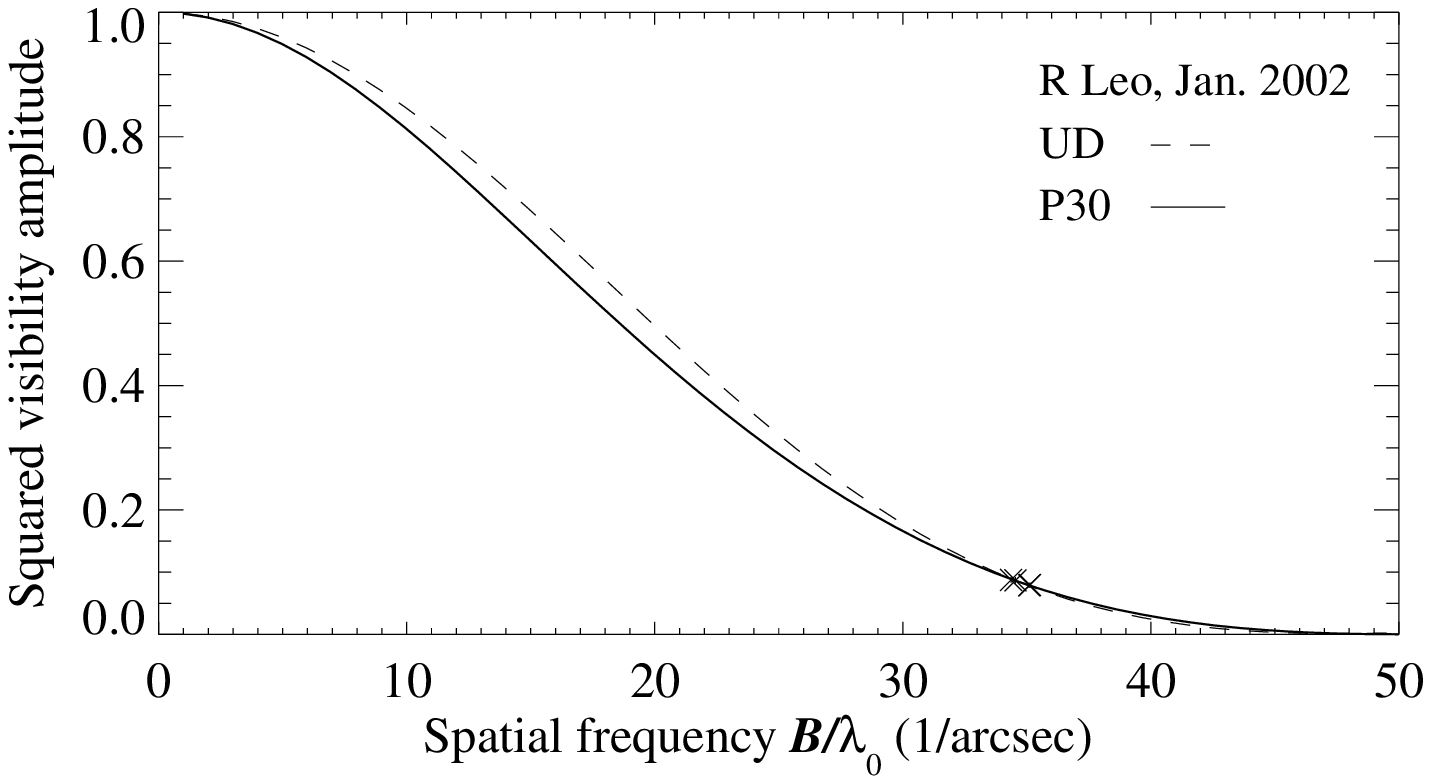}}

\resizebox{\hsize}{!}{\includegraphics{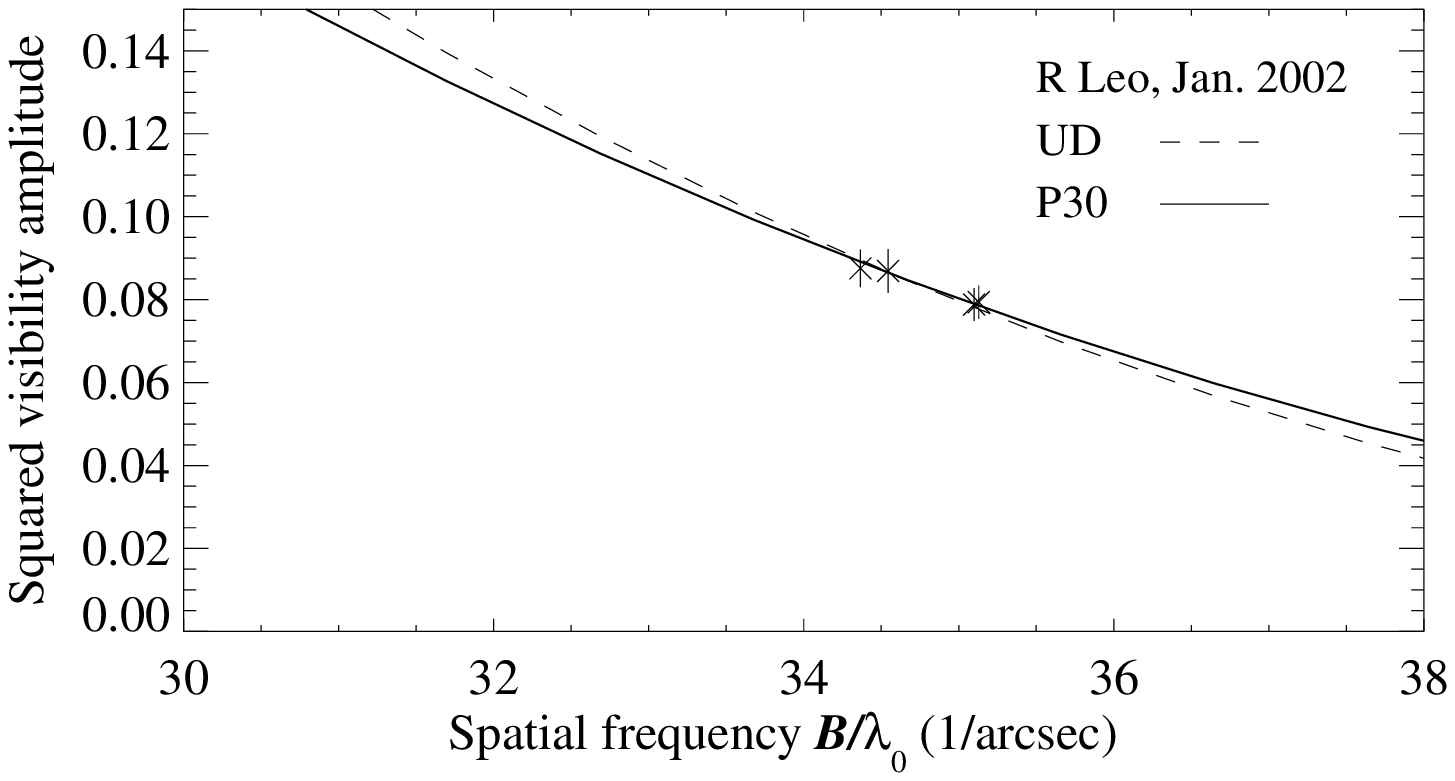}}
\caption{Same as Fig.~\protect\ref{fig:vis2001}, but for 
the January 2002 (visual stellar phase 1.02) data and the P30 model.}
\label{fig:vis2002}
\end{figure}

\begin{table*}
\caption{Fit results to our data using models from the 
P and M series with differences of the visual 
phase between model and observations
($\Delta\phi_\mathrm{vis}=|\phi_\mathrm{Model}-\phi_\mathrm{Obs.}|$) 
up to 0.1 (not taking into account the cycle). Listed are the obtained 
best-fitting Rosseland diameters, and as alternative information 
the 1.04\,$\mu$m continuum diameters, with their formal errors, 
the reduced $\chi^2_\nu$ values, the resulting linear 
Rosseland radii $R_\mathrm{Ross}$ and effective temperatures 
$T_\mathrm{eff}$ obtained with our adopted values for $\pi$ 
and $f_\mathrm{bol}$. The uncertainties of $R_\mathrm{Ross}$ and
$T_\mathrm{eff}$ are $\sim$\,15\% and $\sim$\,10\%, respectively.
The last columns provide a comparison to the 
respective model values of $R_\mathrm{Ross}$ and $T_\mathrm{eff}$.}
\centering
\begin{tabular}{lr|lllll|llll}
\hline\hline
Model & $\Delta\phi_\mathrm{vis}$ & $\Theta_\mathrm{Ross}$ & 
$\sigma(\Theta_\mathrm{Ross})$
& $\Theta_\mathrm{1.04}$ & $\sigma(\Theta_\mathrm{1.04})$ &
$\chi^2_\nu$ & $R_\mathrm{Ross}$ & $T_\mathrm{eff}$ & $R/R_\mathrm{Model}$ &
$T/T_\mathrm{Model}$ \\
     &$=|\phi_\mathrm{Model}-\phi_\mathrm{Obs}|$   & [mas] & [mas] & [mas] & [mas] & & [$R_\odot$] & [K] & & \\
\hline\\[1ex]
\multicolumn{11}{l}{1 \& 3 April 2001, $\phi_\mathrm{vis}=0.08$}\\[1ex]
P10   & -0.08 & 28.9 &  0.03 & 29.1 & 0.03 & 1.86 & 352 & 2906 & 1.42 & 0.93\\
P11n  & +0.02 & 28.7 &  0.01 & 29.2 & 0.00 & 1.56 & 350 & 2916 & 1.25 & 0.98\\
P20   & -0.09 & 27.7 &  0.05 & 27.7 & 0.05 & 0.41 & 337 & 2968 & 1.35 & 0.97\\
P21n  & +0.03 & 28.0 &  0.05 & 27.5 & 0.05 & 0.57 & 341 & 2952 & 1.15 & 1.06\\
P22   & +0.10 & 28.6 &  0.05 & 27.3 & 0.05 & 0.37 & 348 & 2921 & 1.10 & 1.11\\
P30   & -0.10 & 29.0 &  0.04 & 29.3 & 0.04 & 1.29 & 353 & 2901 & 1.30 & 0.95\\
P40   & -0.08 & 28.2 &  0.03 & 27.9 & 0.03 & 0.56 & 344 & 2942 & 1.22 & 1.02\\
M10   & -0.06 & 28.6 &  0.03 & 28.3 & 0.03 & 1.65 & 348 & 2921 & 1.13 & 1.06\\
M11n  & +0.03 & 28.2 &  0.04 & 27.1 & 0.04 & 1.05 & 344 & 2942 & 1.05 & 1.14\\
M20   & -0.03 & 28.6 &  0.04 & 27.9 & 0.03 & 1.32 & 348 & 2921 & 1.09 & 1.10\\
M21n  & +0.02 & 28.6 &  0.04 & 27.5 & 0.03 & 1.26 & 348 & 2921 & 1.06 & 1.15
\\\hline
Avg.      & -0.02 & 28.5 $\pm$ 0.4 & 0.03  & 28.1 $\pm$ 0.8 &
\multicolumn{6}{l}{0.03} \\
\multicolumn{11}{l}{}\\[1ex]
\multicolumn{11}{l}{20 January 2002, $\phi_\mathrm{vis}=1.02$}\\[1ex]
P10   & -0.02 & 26.8 & 0.03 & 27.1 & 0.04 & 0.09 & 327   & 3042 &1.32 & 0.97\\
P11n  & +0.08 & 26.6 & 0.12 & 27.0 & 0.12 & 0.08 & 324   & 3054 &1.15 & 1.02\\
P20   & -0.03 & 24.8 & 0.19 & 24.8 & 0.19 & 0.12 & 303   & 3163 &1.21 & 1.03\\
P21n  & +0.09 & 25.0 & 0.18 & 24.6 & 0.18 & 0.10 & 305   & 3150 &1.03 & 1.13\\
P30   & -0.04 & 26.8 & 0.12 & 27.1 & 0.12 & 0.06 & 327   & 3043 &1.20 & 0.99\\
P40   & -0.02 & 25.9 & 0.13 & 25.6 & 0.13 & 0.03 & 316   & 3095 &1.12 & 1.08\\
M10   & -0.00 & 26.5 & 0.12 & 26.3 & 0.12 & 0.08 & 323   & 3060 &1.04 & 1.11\\
M11n  & +0.09 & 26.1 & 0.11 & 25.0 & 0.11 & 0.05 & 318   & 3083 &0.97 & 1.19\\
M20   & +0.03 & 26.5 & 0.12 & 25.8 & 0.12 & 0.06 & 323   & 3060 &1.01 & 1.15\\
M21n  & +0.08 & 26.5 & 0.12 & 25.4 & 0.11 & 0.06 & 323   & 3060 &0.99 & 1.20\\
\hline
Avg.      & +0.03 & 26.2 $\pm$ 0.7 &0.12 & 25.9 $\pm$ 0.9 &
\multicolumn{6}{l}{0.12}\\
\hline
\end{tabular}
\label{tab:fitresults}
\end{table*}
Our measured R\,Leo squared visibility amplitudes are shown
in Figs.~\ref{fig:vis2001} and \ref{fig:vis2002}, together with a
typical model prediction for each epoch based on the P and M model 
series by Hofmann et al. (\cite{HSW98}), Tej et al. (\cite{tej03}), and 
Ireland et al. (\cite{ISW,ISTW}). A detailed description
of the models and the calculation of synthetic visibility values 
follows below.
The visibility curves
for UD intensity profiles with best fitting UD diameters are shown as well.
The obtained UD diameters and their formal errors for the 
April 2001 and January 2002 data are $\Theta_\mathrm{UD}^K=28.1 \pm 0.05$\,mas
and $\Theta_\mathrm{UD}^K=26.2 \pm 0.01$\,mas, with reduced $\chi^2_\nu$ 
values of 2.47 and 0.14, respectively.
 
The April 2001 data were obtained for a relatively wide range of 
projected baseline lengths (spatial frequencies 
between 31 and 35 cycles/arcsecond) while the azimuth angle of the
projected baseline changed only marginally, thanks to the position of the
star on the sky. This allows the study of the CLV without the often inherent 
simultaneous change of baseline angle and resulting confusion with
possible asymmetries.
These data show that R\,Leo's CLV at this epoch significantly deviates
from a UD intensity profile.
The January 2002 data cover 
only a much smaller range of spatial frequencies between 34 and 35 
cycles/arcsecond, and do not allow a discrimination of UD and more complex
intensity profiles.

The P and M model series are complete self-excited dynamic model
atmospheres that were created to match the parameters
of the best studied Mira variables $o$\,Cet and R\,Leo. 
These series are based on fundamental mode pulsation. 
First-overtone pulsation models, such 
as the O model series of Hofmann et al. (\cite{HSW98})
are not considered since recent results indicate that Mira stars
are fundamental mode pulsators (see Introduction). For the
details of the model calculations, including assumptions and 
approximations, we refer to Hofmann et al. (\cite{HSW98}).
The P and M series differ with respect to
the mass of the so-called ``parent star'', which is a 
hypothetical non-pulsating giant that has the same mass and luminosity 
as the Mira variable. The geometric
pulsation of the Mira occurs around the parent's star Rosseland radius
$R = R(\tau_\mathrm{Ross}=1)$ (e.g. Ireland et al. \cite{ISW,ISTW}). 
The parent star of the here considered P and M series has 
solar metallicity, luminosity
$L/\mathrm{L}_\odot$=3470, period 332 days,
mass $M/\mathrm{M}_\odot$=1.0 (P series) and 1.2 (M),
radius $R/\mathrm{R}_\odot$=241 (P) and 260 (M)
(Hofmann et al. \cite{HSW98}).
The moderately 
larger radius of the M-series parent star leads to a slightly lower effective 
temperature of the parent star and systematically lower phase-dependent 
effective temperatures of the pulsating Mira compared to the P series (see 
the discussion in Ireland et al. \cite{ISTW}). The M models tend to exhibit 
less pronounced cycle-to-cycle variations than the P models and to have 
more compact
atmospheres resulting in only moderate differences of K-band contamination but 
larger differences of high-layer absorption features 
(cf. Hofmann et al. \cite{HSW98}; Ireland et al. \cite{ISW,ISTW}).

\begin{figure}
\resizebox{\hsize}{!}{\includegraphics{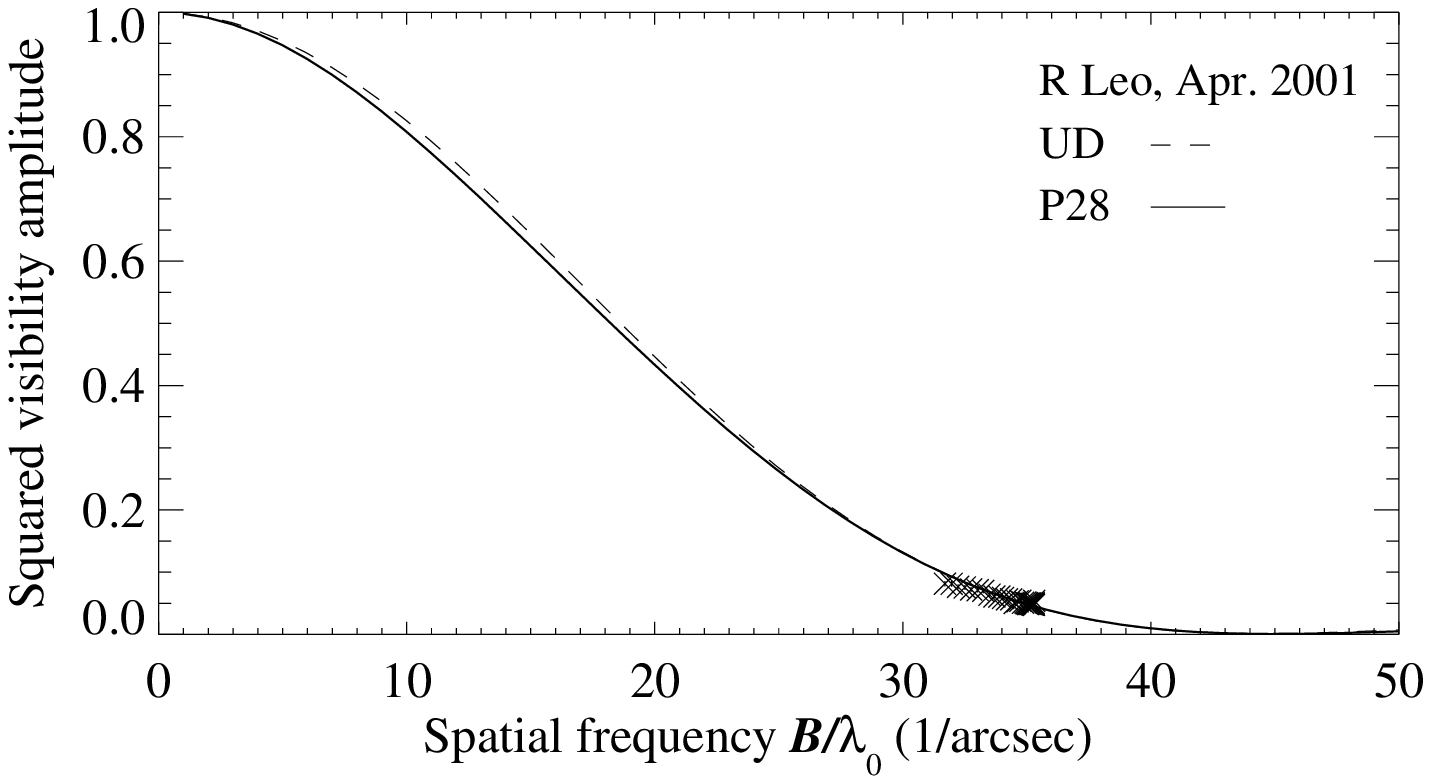}}

\resizebox{\hsize}{!}{\includegraphics{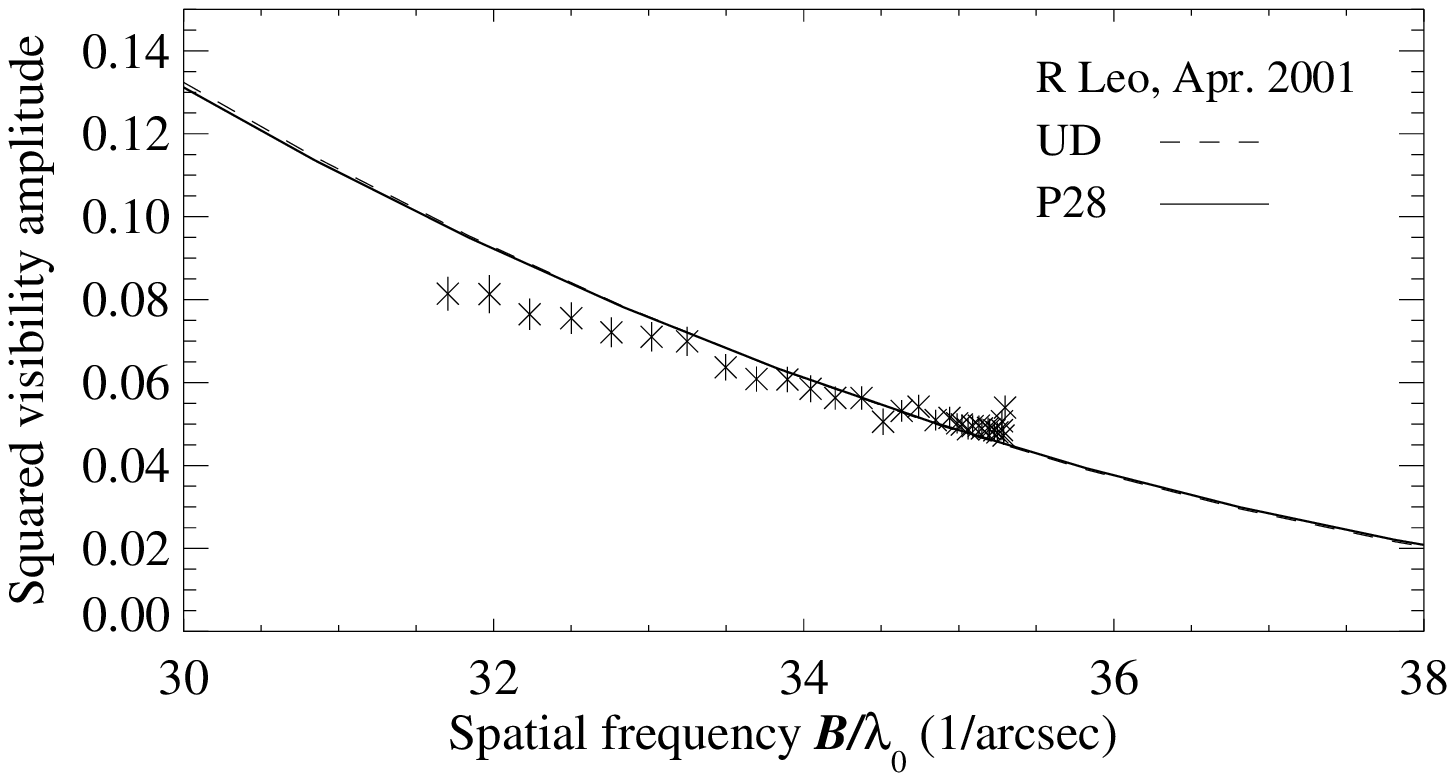}}
\caption{For illustration, measured R\,Leo squared visibility
amplitudes obtained in April 2001 (visual stellar phase 0.08),
as in Fig.~\protect\ref{fig:vis2001}, but together
with a typical pre-maximum (P28) model prediction (visual phase 0.83), 
which is as expected not consistent with our
post-maximum measurement ($\chi^2_\nu$=2.47). The model-predicted
CLV of this pre-maximum model is very close to a UD model because
it includes mainly the CLVs of continuum-forming layers without
much molecular contamination. Consistently, a near-UD $K$-band visibility 
curve was observed for R Leo by Monnier et al. (\protect\cite{monnier})
at pre-maximum phase 0.71.}
\label{fig:visP28}
\end{figure}

These models are available for 25 phase and cycle combinations
for the P series (see Ireland et al. \cite{ISW}), and for
20 combinations for the M series (see Ireland et al. \cite{ISTW}).
The models include molecular shells close to
continuum-forming layers. The model studies by Ireland et al. 
(\cite{ISW}, \cite{ISTW}) show that the true continuum CLVs are relatively
close to UD profiles and follow a sinusoidal diameter variation
as a function of photometric stellar phase. For observations with
broad near-continuum bandpasses in the near-IR that include molecular 
(mainly water) lines, 
the molecular shells lead to tail-like or protrusion-like extensions of 
the CLV and to deviations from the sinusoidal diameter variation
(cf. Tej et al. \cite{tej03a}).
This effect due to molecular shells is smallest at pre-maximum phases 
($\sim$ 0.8) of the photometric variability when continuum photons 
are mostly generated in a narrow region just below the strong 
shock emerging from the stellar interior. At these phases,
effective temperatures are highest, and the visibility
curves for broad bandpasses are expected to be very close to
UD curves up to the first minimum (see the example in Fig.~\ref{fig:visP28}). 
For maximum and post-maximum phases, however, the molecular contamination
is higher, the CLVs for broad bandpasses 
are clearly different from UD profiles,
and these differences can already be detected in the first lobe of the
visibility function (see the examples in Figs.~\ref{fig:vis2001} and 
\ref{fig:vis2002} below).

For each model, we used monochromatic tabulated intensity values at 
46 radii between 0 and 5 parent star radii $R_p$, for wavelengths
between 1.99 and 2.40 $\mu$m in steps of 0.01$\mu$m. Broad-band
synthetic visibility values for the VINCI sensitivity function
($\sim$\,1.9--2.5\,$\mu$m) were computed as described in 
Wittkowski et al. (\cite{WAK:04}). The VINCI sensitivity curve
includes the transmission curves of the VINCI $K$-filter, the sensitivity
of the optical fibers, the detector quantum efficiency, and the 
atmospheric transmission.
The wings of the VINCI sensitivity function between 
1.9 and 2.0\,$\mu$m, as well as between 2.4 and 2.5\,$\mu$m are 
neglected. A dust shell was not included in the modeling.
The dust emission coefficient 
of 0.24 (Sloan \& Price \cite{sloan}, see Sect. \ref{sec:introduction})
is relatively low, and dust shells as optically thin 
as this, are not expected to contribute significantly to $K$-band 
visibilities (see e.g., Woodruff et al. \cite{woodruff}, 
Monnier et al. \cite{monnier}, Ohnaka et al. \cite{ohnaka}; see also
Bedding et al. \cite{bedding}).
Since all observations cover only a narrow range of azimuth angles of 
the projected baseline, our observations are not
sensitive to asymmetries (see above and Tab.~\ref{tab:obs}). As a result,
asymmetric shapes of the intensity distribution are not considered for 
the modeling.

We estimate the uncertainty of the assignment of the model stellar 
variability phases relative to the stellar phases at the dates of our 
observations to be about 0.1 owing to irregularities and cycle-to-cycle 
variations of the light curves of modeled and real stars 
(cf. Ireland et al. \cite{ISW,ISTW}). As a result,
fits were made to our observational data with all models that have
a (model) phase which differs by up to 
0.1 from the stellar phases at the dates of our observations.
For each model, the angular diameter is the only free parameter of the fit. 
In order to characterize the angular diameter of the fitted CLV,
any well-defined reference radius of the model CLV can be used, such
as the Rosseland radius or the 1.04\,$\mu$m continuum radius.
Physically most meaningful may be a true continuum radius, such as
the 1.04\,$\mu$m radius, which is not affected by time variable
molecular contamination (see Hofmann et al. \cite{HSW98}; Jacob \& Scholz
\cite{JS02}; Ireland et al. \cite{ISW,ISTW}). In the following, 
the Rosseland 
radius is mainly used as reference quantity, as is usual in the literature. 
It can be transformed into any other reference point of the model CLVs
with conversion factors derived from the model atmospheric structure,
as for instance those given in Ireland et al. (\cite{ISW,ISTW}).

Tab.~\ref{tab:fitresults} lists for these models the
best-fitting Rosseland diameters, and as alternative information,
the 1.04\,$\mu$m continuum diameters, with their formal errors and 
the reduced $\chi^2_\nu$ values. Shown are also the resulting linear 
Rosseland radii $R_\mathrm{Ross}$ and the effective temperatures 
$T_\mathrm{eff}$ computed with our adopted values for the parallax $\pi$ 
and the bolometric fluxes $f_\mathrm{bol}$ (see above). The last 
columns provide a comparison to the respective model values 
of $R_\mathrm{Ross}$ and $T_\mathrm{eff}$. Our derived
$R_\mathrm{Ross}$ values have an uncertainty of $\sim$\,15\%, and 
the $T_\mathrm{eff}$ values of $\sim$\,10\%. 
Taking into account the uncertainties of our visibility values and 
of the derived $R_\mathrm{Ross}$ and $T_\mathrm{eff}$ values, all 
considered model predictions are consistent with our measured 
visibility values and with the derived $R_\mathrm{Ross}$ 
and $T_\mathrm{eff}$ values, to within 1-2$\sigma$. 
The correspondence of our obtained linear radii with the
model radii also indicates that the assumption of fundamental mode
pulsation is valid. 
For instance, the (piston-driven)
first-overtone E model series of Bessell et al. (\cite{bessell}) predict 
radii (given in Hofmann et al. \cite{HSW98}) which are 
larger by a factor of 1.5 when compared to the fundamental mode
P series and by a factor of  1.4 when compared to the M series.
The (self-excited) overtone O series of Hofmann et al. (\cite{HSW98}) 
have radii larger by factors 2.1 and 1.9, respectively.
The mean $\Theta_\mathrm{Ross}$ values, 
averaged over the fits to all considered models, are used as the final
angular diameters. 
The average formal error of $\Theta_\mathrm{Ross}$ 
is $\pm$\,0.03\,mas for the April 2001 data, and $\pm$\,0.12\,mas for the 
January 2002 data. In addition, a calibration error of $\pm$\,0.1\,mas 
is considered, based on calibrations using different calibrators, and 
different averaging of the transfer function over the night. Finally, 
an uncertainty arises from the scatter of the fit results using the
different
models, which amounts to $\pm$\,0.39\,mas for the April 2001 data, and
$\pm$\,0.72\,mas for the January 2002 data. Altogether, the final
values for the Rosseland angular diameters are\\[1ex]
$\Theta_\mathrm{Ross}(\mathrm{April~2001,~phase~ 0.08})
=28.5\pm0.4$\,mas and\\
$\Theta_\mathrm{Ross}(\mathrm{January~2002,~phase~ 1.02})
=26.2\pm0.8$\,mas.\\[1ex]
The Rosseland angular diameter at the variability phase closer to the
maximum (1.02) is smaller by $\sim$\,8\,\% than that at variability
phase 0.08. This is consistent with pulsation models 
(see, e.g. Ireland et al. \cite{ISW,ISTW}).
Together with the adopted values for $\pi$ and $f_\mathrm{bol}$, these
angular diameters for  April 2001 and January 2002 correspond to
linear radii of $350^{+50}_{-40}$\,R$_\odot$ and 
$320^{+50}_{-40}$\,R$_\odot$, and to effective temperatures 
of $2930 \pm 270$\,K and $3080 \pm 310$\,K, respectively.
\begin{figure}
\resizebox{0.9\hsize}{!}{\includegraphics{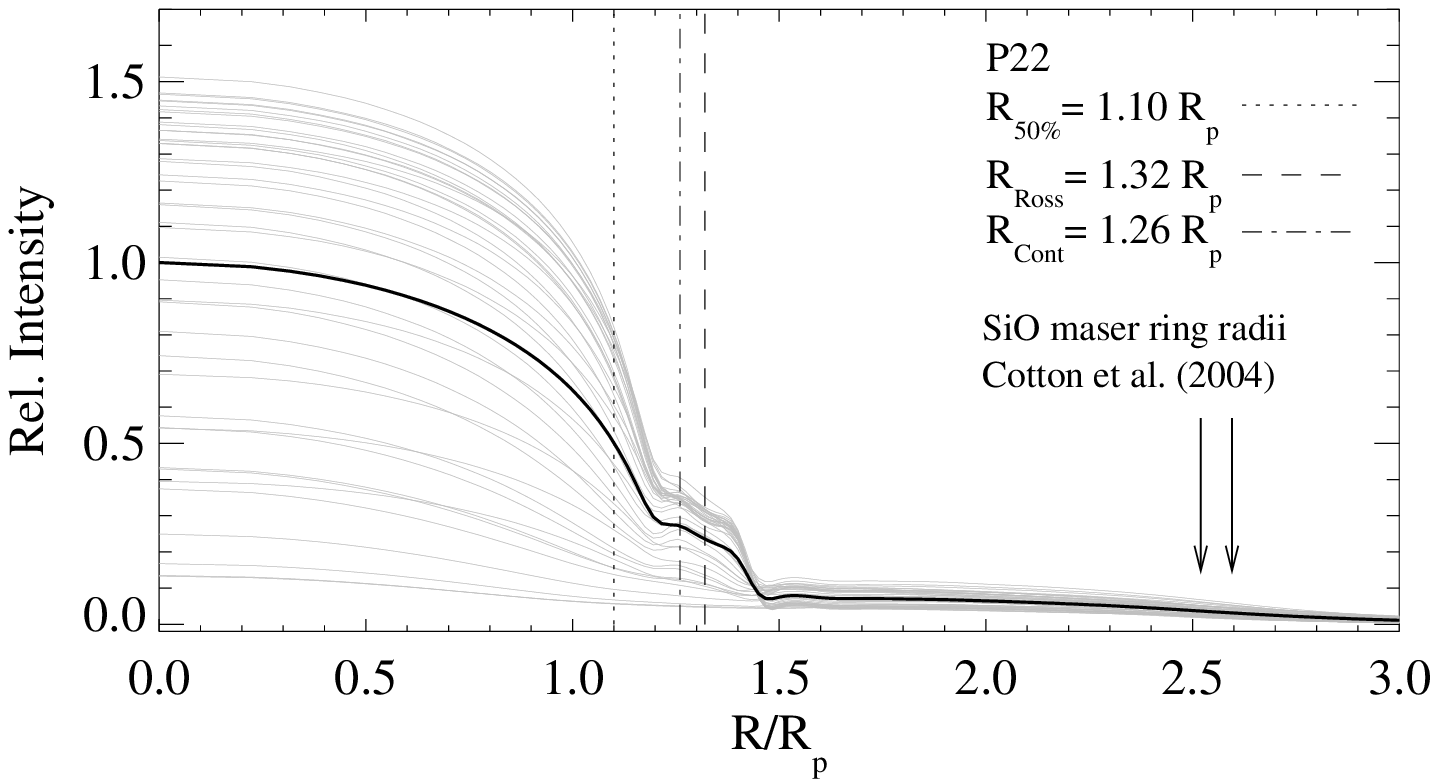}}

\resizebox{0.9\hsize}{!}{\includegraphics{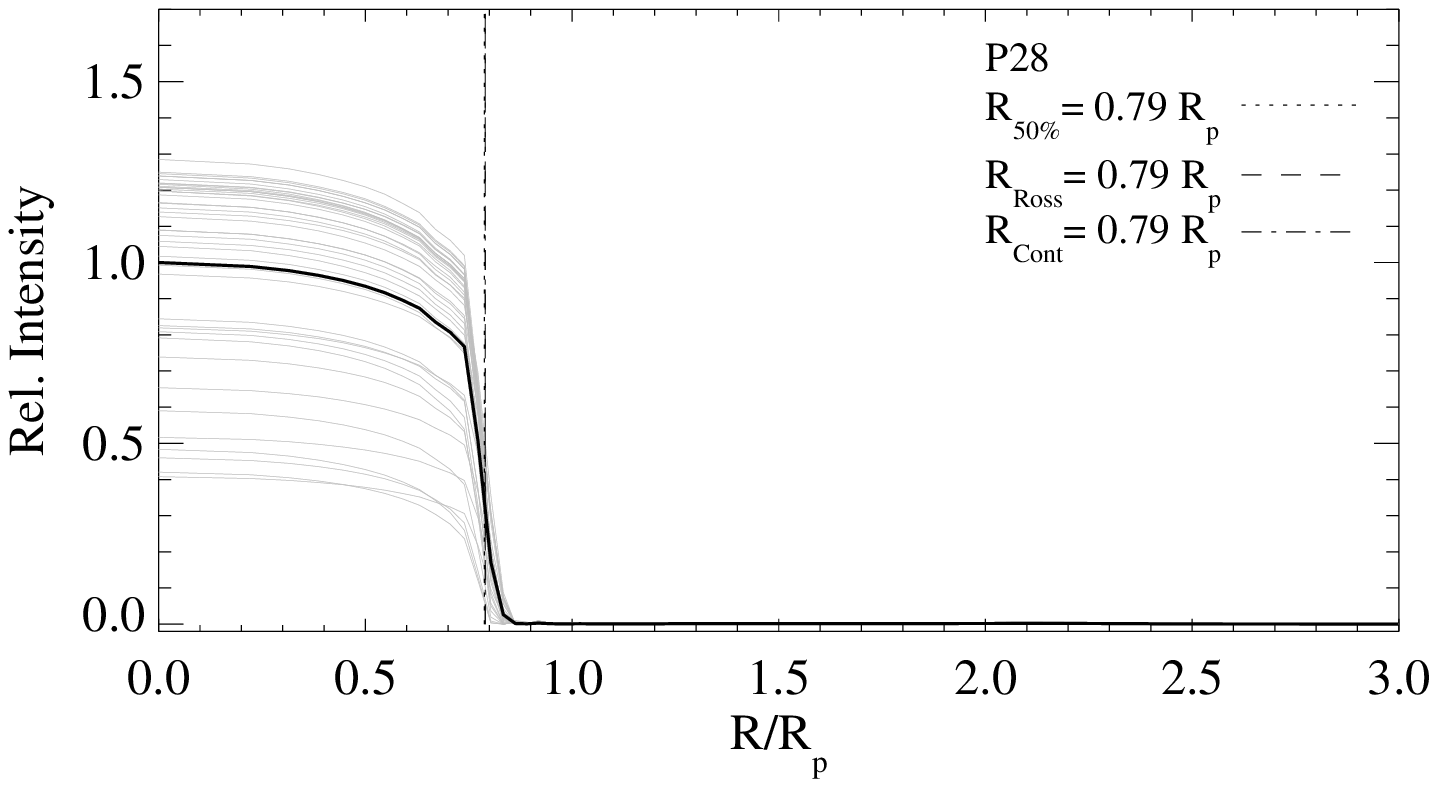}}

\resizebox{0.9\hsize}{!}{\includegraphics{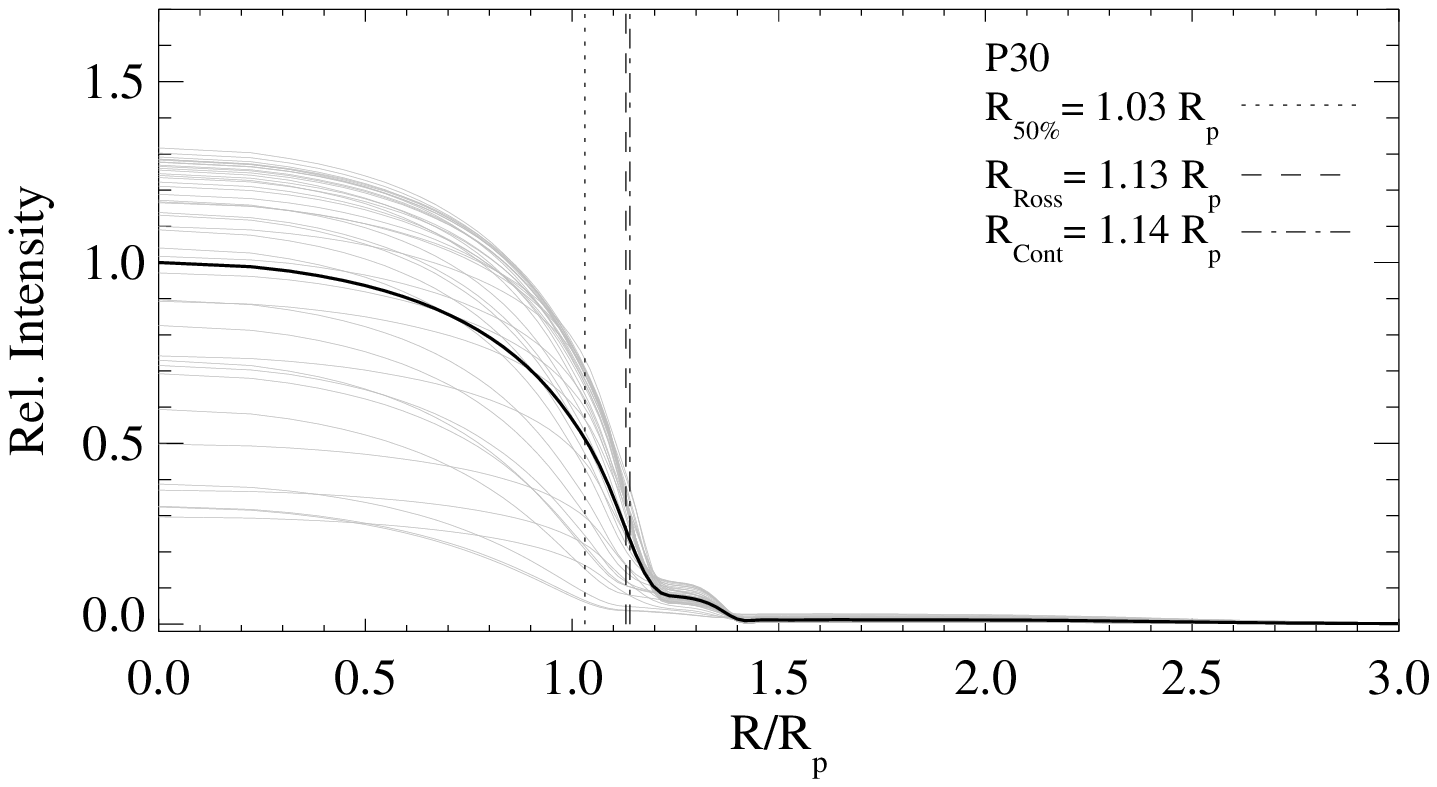}}
\caption{CLV predictions from the (top) P22, (middle) P28,
and (bottom) P30 model, corresponding to the model visibility
curves in Figs.~\ref{fig:vis2001}, \ref{fig:visP28}, and
\ref{fig:vis2002}, respectively. The P22 and P30 models fit well
our April 2001 and January 2002 data, respectively. The P28 model,
a typical pre-maximum model, is as expected not consistent with
our post-maximum measurements.
The thin lines denote
the monochromatic CLVs while the thick line denotes the 
CLV averaged over the VINCI sensitivity function. Indicated are
also (dashed line) the Rosseland radius, (dashed-dotted line) the
1.04\,$\mu$m true continuum radius, as well as (dotted line) the 
radius at which the filter-averaged CLV drops by 50\%.
The mean SiO maser ring radii (for the 43.1\,GHz and 42.8\,GHz transitions)
measured by Cotton et al. (\protect\cite{cotton}) close in time
to our April 2001 data are indicated by the arrows in the top panel.
}
\label{fig:clv}
\end{figure}

The $\chi^2_\nu$ values in Tab.~\ref{tab:fitresults} suggest
that the models of the second cycle of the P series (P2x models)
fit  our April 2001 data marginally better than those of the other model cycles.
However, these differences are not significant, and a better baseline 
coverage would be needed to clearly discriminate between different 
model cycles and between the P and M model series 
(see the discussion in Ireland et al. \cite{ISW,ISTW}).
As an example, Figs.~\ref{fig:vis2001} 
and \ref{fig:vis2002} show comparisons of the April 2001 data to the 
well fitting P22 model, and of the January 2002 data to the
P30 model prediction, respectively. These models match the data well 
at the respective stellar phases, in terms of consistency of the CLV and
of the $T_\mathrm{eff}$ and $R_\mathrm{Ross}$ values. For comparison, 
best fitting UD curves are shown as well. Fig.~\ref{fig:visP28} shows
for illustration for our April 2001 data the VINCI bandpass-averaged
visibility prediction for a typical 
{\it pre-maximum} model, 
P28 (stellar phase 0.83), which is very close to a UD profile, 
and as expected not consistent with our {\it post-maximum} measurement.

Fig.~\ref{fig:clv} shows the corresponding
P22, P28, and P30 CLVs. Plotted are the monochromatic 
CLVs as well as the filter-averaged CLV. The Rosseland- and the
1.04\,$\mu$m true continuum radii are indicated, as well as the 
radius at which the filter-averaged CLV decreases by 50\%.
For the near-UD pre-maximum CLV of P28, these different radii coincide
because there is little molecular contamination. 
For the maximum and
post-maximum models P22 and P30 these three different radius definitions
for each model differ by 15\% and 10\%, respectively, owing to a mix of 
contributions from continuum and line-forming layers in our broad bandpass
at these phases of the pulsation cycle.

\section{Comparison to other observations}
Our R\,Leo angular diameter values are consistent with earlier $K$-band 
measurements listed in Tab.\ref{tab:diams}, when taking into account
possible phase and cycle-to-cycle variations.

Very recently, Perrin et al. (\cite{perrin2004}) presented 
high-precision interferometric measurements of R\,Leo in four 
relatively narrow sub-filters
of the $K$-band obtained in November 2001 at variability phase 0.80. 
They compare 
these measurements together with $L$ broad-band observations from 2000 at
phase 0.63 to an empirical model. This model consists of a stellar uniform disk
and a single thin molecular layer detached from the star with different
optical depths at each observed wavelength (9 free parameters), and leads to
a good agreement with the data. They obtain a stellar (continuum) diameter of
21.88\,$\pm$\,1.7\,mas at their pre-maximum variability phase. This diameter
estimate is consistent with our values of the continuum diameter of 
28.1\,$\pm$\,0.8\,mas and 25.9\,$\pm$\,0.9\,mas at variability phases 0.08 and
1.02, considering that the P and M model series predict diameter differences 
between phases 0.8 and 1.1 by a factor of about 1.5 (P08 compared to P11n; 
P17n to P21n; M08 to M11n; M18 to M21n; see
Ireland et al. \cite{ISW,ISTW}). It has to be kept in mind as well
that the modeling approaches are quite different and that the assignment 
of the model phase to the observational phase
has some uncertainty (see Sect.~\ref{sec:models}). 
In the here adopted 
approach, the fundamental stellar parameters are the only free parameters and
the CLV in any bandpass is then predicted as a function of phase. The 
measurements of Perrin et al. (\cite{perrin2004}) also
illustrate the importance of narrow-band observations for more
detailed tests of Mira atmosphere models.

Woodruff et al. (\cite{woodruff}) recently measured the shape of the
$K$-band visibility curve for \object{$o$\,Cet} at post-maximum stellar phase
0.13 and found a similar deviation from a UD curve as for our post-maximum
R\,Leo observation at phase 0.08, both of which are consistent with
the predictions of the P model series at such stellar phases. This
similarity between the R\,Leo and $o$\,Cet CLVs at similar phase
is remarkable since the shapes and amplitudes of their lightcurves
are different (Ireland et al. \cite{ISW,ISTW}).  

Monnier et al. (\cite{monnier}) report that their measured R\,Leo 
$K$-band visibility function at pre-maximum phase 0.71 
(around February 1, 2000)
can be well be described by a UD curve.
This is consistent with the 
predictions from the P and M model series which show near-UD CLVs for broad
bandpasses at pre-maximum stellar phases (see Sect.~\ref{sec:models}).
Consistently, Perrin et al. (\cite{perrin}) measured a R\,Leo $K$-band 
visibility function which is clearly different from a UD curve at post-maximum 
stellar phase $\sim$\,0.28. 

Cotton et al. (\cite{cotton}) have obtained VLBA maps of
the $v=1, J=1-0$ (43.1 GHz) and $v=2, J=1-0$ (42.8 GHz)
SiO maser emission toward several Mira stars, including R\,Leo.
One of their R\,Leo epochs, from 29 April 2001, is very close
in time to our first VLTI epoch (1 \& 3 April 2001), with a difference
of the stellar phase of only 0.09. They derived a mean diameter of the
SiO maser rings of 51.1\,mas and 49.3\,mas for the $v=1$ and $v=2$ 
transitions, respectively (with widths of 2.2\,mas). 
Compared to our value for the April 2001 R\,Leo Rosseland angular diameter,
the SiO maser spots for the $v=1$ and $v=2$ transitions
lie at $\sim$\,1.8 and $\sim$\,1.7 Rosseland radii, respectively.
Cotton et al. (\cite{cotton}) obtained slightly
larger mean $R_\mathrm{SiO}/R_\mathrm{2.2\,\mu m}$
values of 2.0 and 1.9, respectively,  
when comparing to a mean value of the near-infrared $K$-band UD diameters 
from Perrin et al. (\cite{perrin}) obtained in April 1996 and March 1997.
Our newly estimated R\,Leo $R_\mathrm{SiO}/R_\mathrm{Ross}$ values
are also in good agreement 
with recent $R_\mathrm{SiO}/R_\mathrm{Ross}$ values of
$\sim$\,1.7 and $\sim$\,1.6 for the same $\nu=1$ and $\nu=2$ SiO 
transitions obtained by Boboltz \& Wittkowski (\cite{BW}) for the
Mira star \object{S\,Ori} by quasi-simultaneous VLTI/VLBA observations. 
For illustration, the mean maser ring radii derived by 
Cotton et al. (\cite{cotton}) are indicated in the top panel of 
Fig.~\ref{fig:clv}, together with the well fitting P22 near-infrared 
$K$-band model CLV.

Danchi et al. (\cite{danchi}) obtained an inner radius of the dust envelope
of R\,Leo of 70\,mas by radiative transfer modeling using mid-infrared
interferometric data taken in 1988 and 1990. Compared to our Rosseland
angular diameter, this would place the inner radius of the dust envelope
at roughly 5 Rosseland radii, and at roughly three times the distance
of the SiO maser zone.
\section{Summary and conclusions}
We have obtained near-infrared $K$-band interferometric data on R\,Leo
for two near-maximum (visual stellar phases 0.08 and 1.02) epochs. 

For the first epoch, our interferometric data -- obtained for
a sufficiently large range of spatial frequencies --
show that R\,Leo's intensity profile deviates significantly from a 
uniform disk profile. We have compared these data to recent 
self-excited dynamic
models which were specially created to match the stellar parameters 
of $o$\,Cet and R\,Leo. We find that these model CLVs for the phases
of our observations are consistent with our measurements.
Furthermore, the model parameters for $R_\mathrm{Ross}$ and 
$T_\mathrm{eff}$ are consistent with those derived from our 
observations, as well. The models include the effects of close molecular 
layers, and the correspondence with our measurements show that
these effects are well described. 

Very recently, Woodruff et al. (\cite{woodruff}) found a similar shape of 
the $K$-band visibility curve for the Mira star $o$\,Cet at post-maximum 
phase 0.13, and consistency with predictions by the P-model 
series (model P21n). It is remarkable that the two stars appear to show 
similar CLVs which are consistent with the predictions by the 
P model series, while their lightcurves  and variability amplitudes are 
different (Ireland et al. \cite{ISW,ISTW}).
In accordance with the CLV predictions by the P and M model series,
Perrin et al. (\cite{perrin}) also obtained a R\,Leo visibility function
in the $K$-band which is very different from a UD curve at post-maximum 
stellar phase 0.28, and Monnier et al. (\cite{monnier}) report a 
$K$-band visibility function which can be well be described by a UD curve
at pre-maximum stellar phase 0.71.  
These findings increase our confidence in these dynamic Mira star
models, which are often used to transform broad-band filter-specific
UD diameters into more meaningful Rosseland or continuum diameters.

We also obtained high-precision angular Rosseland diameters for R\,Leo
at the epochs of our observations, and derive fundamental parameters, 
i.e. linear radius and effective temperature, 
from these in combination with
literature estimates of R\,Leo's distance and bolometric flux.
The correspondence of our obtained linear radii with model radii
of the fundamental mode pulsation models used is in agreement
with the general recent conclusions
that Mira stars pulsate in fundamental mode.

More detailed observations are desirable in the future in order
to better constrain the models. Such observations should
probe the CLV at a larger range of spatial frequencies. In addition,
measurements with high spectral resolution 
in both true continuum and certain molecular bands will be useful in 
order to separate line-forming and continuum-forming layers. 
Moreover, monitoring of the observed CLVs in time over several cycles
with a resolution of $\sim$\,10\% of the variability period are
desirable in order to investigate the strong model-predicted 
CLV variations with variability phase and cycle.
\begin{acknowledgements}
We thank D.~Boboltz, T.~Driebe, and K.~Ohnaka for interesting discussions, 
and I.~Percheron and V.~Roccatagliata for help and support with the
data reduction. We are grateful for the valuable comments on our manuscript 
provided by the referee B.~Lopez.
DF's stay at ESO was partly financed by the Italian National Institute for
Astrophysics (INAF), under financial fund no. 0330909. 
We are also grateful for partial support by the ESO DGDF 2004.
This research was in part supported by the Australian Research Council and
the Deutsche Forschungsgemeinschaft within the linkage project ``Red Giants''.
We acknowledge with thanks the variable star observations from the AAVSO 
International Database contributed by observers worldwide and used in this 
research.
This research has made use of the AFOEV database, operated at CDS, France.
\end{acknowledgements}


\begin{thebibliography}{}
%
\bibitem[2001]{bedding}
Bedding, T.~R., Jacob, A.~P., Scholz, M., \& Wood, P.~R.\ 2001, \mnras, 
325, 1487
%
\bibitem[1996]{bessell}
Bessell, M.~S., Scholz M., \& Wood, P.~R.\ 1996, \aap, 307, 481
%
\bibitem[2004]{BW}
Boboltz, D.~A., \& Wittkowski, M.\ 2005, \apj, in press
%
\bibitem[2002]{borde}
Bord{\'e}, P., Coud{\'e} du Foresto, 
V., Chagnon, G., \& Perrin, G.\ 2002, \aap, 393, 183
%
\bibitem[1998]{burns} 
Burns, D., Baldwin, J.~E., Boysen, R.~C., et al.\ 1998, \mnras, 297, 462
%
\bibitem[2004]{cotton}
Cotton, W.~D., Mennesson, B., Diamond, P.~J., et al.\ 2004, \aap, 414, 275
%
\bibitem[1994]{danchi} 
Danchi, W.~C., Bester, M., Degiacomi, C.~G., Greenhill, L.~J., \& Townes, 
C.~H.\ 1994, \aj, 107, 1469 
%
\bibitem[1991]{digiacomo} 
di Giacomo, A., Lisi, F., Calamai, G., \& Richichi, A.\ 1991, \aap, 249, 397 
%
\bibitem[1996]{dyck}
Dyck, H.~M., Benson, J.~A., van Belle, G.~T., \& Ridgway, S.~T.\ 1996, \aj, 
111, 1705 
%
\bibitem[1992]{gatewood}
Gatewood, G.\ 1992, \pasp, 104, 23
%
\bibitem[1995]{haniff}
Haniff, C. A., Scholz, M., \& Tuthill, P. G. 1995, \mnras, 276, 640
%
\bibitem[1998]{HSW98}  
Hofmann, K.-H., Scholz, M., \&  Wood, P.R.\ 1998, \aap, 339, 846
%
\bibitem[2001]{hofmann}
Hofmann, K.-H., Balega, Y., Scholz, M., \& Weigelt, G.\ 2001, \aap, 376, 518
%
\bibitem[2002]{hofmann02}
Hofmann, K.-H., Beckmann, U., Bl\"ocker, T., et al.\ 2002, \na, 7, 9
%
\bibitem[2004a]{ISW} 
Ireland, M. J., Scholz, M., Wood, P. R.\ 2004a, \mnras, 352, 318
%
\bibitem[2004b]{ISTW}
Ireland, M.~J., Scholz, M., Tuthill, P.~G., \& Wood. P.~R.\ 2004b, \mnras,
in press
%
\bibitem[2002]{JS02} 
Jacob, A.~P.~\& Scholz, M.\ 2002, \mnras, 336, 1377 
%
\bibitem[2003]{kervella1}
Kervella, P., Gitton, P., \& S{\'e}gransan, D., et al.\ 2003, 
\procspie, 4838, 858
%
\bibitem[2004]{kervella2} 
Kervella, P., S{\'e}gransan, D. \& Coud{\'e} du Foresto, V.\ 2004,
\aap, 425, 1161
%
\bibitem[1998]{knapp} 
Knapp, G.~R., Young, K., Lee, E., \& Jorissen, A.\ 1998, \apjs, 117, 209
%
\bibitem[1998]{kholopov}
Kholopov, P.~N., et al.\ 1998, Combined General Catalogue of 
Variable Stars, 4.1 Ed (II/214A).~(1998)
%
\bibitem[2002]{mennesson}
Mennesson, B., Perrin, G., Chagnon, G., et al. 2002, \apj, 579, 446
%
\bibitem[2004]{monnier} 
Monnier, J.~D., Millan-Gabet, R., Tuthill, P.~G., et al.\ 2004, \apj, 605, 436
%
\bibitem[2004]{ohnaka}
Ohnaka, K., Bergeat, J., Driebe, T., et al. \ 2004, \aap, in press
%
\bibitem[1999]{perrin} 
Perrin, G., Coud{\'e} du Foresto, V., Ridgway, S.T., et al.\ 1999, \aap, 345, 221
%
\bibitem[2004]{perrin2004}
Perrin, G., Ridgway, S.T., Mennesson, B., et al.\ 2004, \aap, 426, 279
%
\bibitem[1997]{esa} 
Perryman, M.~A.~C.~\& 
ESA 1997, The Hipparcos and Tycho catalogues, 
Publisher: Noordwijk, Netherlands: ESA Publications Division, 1997, Series: 
ESA SP Series vol no: 1200, ISBN: 9290923997
%
\bibitem[1998]{sloan}
Sloan, G.~C.~\& Price, S.~D.\ 1998, \apjs, 119, 141
%
\bibitem[1998]{Scholz98} 
Scholz, M.\ 1998, IAU Symp.~189: Fundamental Stellar Properties, 189, 51 
%
\bibitem[2001]{Scholz01} 
Scholz, M.\ 2001, \mnras, 321, 347
%
\bibitem[2003]{Scholz03} 
Scholz, M.\ 2003, \procspie, 4838, 163 
%
\bibitem[2000]{scholzwood}
Scholz M., Wood P.R.\ 2000, \aap, 362, 1065
%
\bibitem[1999]{tej99} 
Tej, A., Chandrasekhar, T., Ashok, N.~M., et al.\ 1999, \aj, 117, 1857
%
\bibitem[2003a]{tej03a}
Tej, A., Lan{\c c}on, A., \& Scholz, M.\ 2003a, \aap, 401, 347 
%
\bibitem[2003b]{tej03} 
Tej, A., Lan{\c c}on, A., Scholz, M., \& Wood, P.~R.\ 2003b, \aap, 412, 481
%
\bibitem[1996]{vanbelle1}
van Belle, G.~T., Dyck, H.~M., Benson, J.~A., \& Lacasse, M.~G.\ 1996, \aj, 
112, 2147
%
\bibitem[2002]{vanbelle2}
van Belle, G.~T., Thompson, R.~R., \& Creech-Eakman, M.~J.\ 2002, \aj, 124, 
1706
%
\bibitem[2004]{AAVSO} Waagen, E. O.\ 2004, Observations from the AAVSO 
International Database, private communication
%
\bibitem[2000]{WF} 
Whitelock, P.~\& Feast, M.\ 2000, \mnras, 319, 759
%
\bibitem[2000]{whitelock}
Whitelock, P., Marang, F., \& Feast, M.\ 2000, \mnras, 319, 728
%
\bibitem[2001]{WHJ:01}
Wittkowski, M., Hummel, C.~A., Johnston, K.~J., et al.\ 2001, \aap, 377, 981
%
\bibitem[2004]{WAK:04} 
Wittkowski, M., Aufdenberg, J.~P., \& Kervella, P.\ 2004, \aap, 413, 711
%
\bibitem[1999]{wood}
Wood, P.~R., Alcock, C., Allsman, R.A., et al.\ 1999, 
IAU Symp.~191: Asymptotic Giant Branch Stars, 191, 151
%
\bibitem[2004]{woodruff}
Woodruff, H.~C., Eberhardt, M., Driebe, T., et al.\ 2004, \aap, 421, 703
%
\bibitem[2000]{young}
Young, J.~S., Baldwin, J. E., Boysen, R.~C., et al.\ 2000, \mnras, 318, 381 
\end{thebibliography}
\end{document}